# Defect solitons in defective PT potentials with real part of dual-frequency lattices

Yunji Meng , Youwen Liu[*], Peining Li

*Department of Applied Physics, Nanjing University of Aeronautics and Astronautics, Nanjing 210016, China*

*Corresponding author: ywliu@nuaa.edu.cn*

**Abstract**

We address the properties of defect solitons (DSs) in defective parity-time potentials with real part of dual-frequency lattices. The impact of defect on stability regions of DSs was considered. For positive defects, DSs whose real/imaginary parts are symmetric/antisymmetric (SA) functions of position are always stable in the semi-infinite gap and nonexistent in the first gap. While for negative defects, in semi-infinite gap, the SADSs are stable in most of their existence region besides low power region, but in the first gap, all the SADSs are stable. The ASDSs are unstable in the whole semi-infinite gap regardless of defects, but can be stable in the low power region for positive defects.

## 1. Introduction

The Dirac-von Neumann formulation of quantum mechanics demands that every physical observable is related to a real spectrum and thus the Hamiltonian of corresponding physical system must be Hermitian [1]. The last requirement not only implies real energy egienvalues but also ensures a unitary temporal evolution. Interestingly, a decade ago, C. M. Bender *et al* demonstrated that even non-Hermitian Hamiltonians can exhibit an entirely real spectrum provided they respect parity-time (PT) symmetry [2-5]. A necessary (but not sufficient) condition for a Hamiltonian to be PT symmetric is that its potential $R(x)$ satisfies the condition $R(x) = R^*(-x)$, that is to say the real part of the complex potential involved must be a symmetric function of position, while the imaginary component should be antisymmetric. Due to their unique nature that are absent in Hermitian Hamiltonian, PT-symmetric Hamiltonians have inspired discussions on several fronts in physics, including quantum field theories [6], non-Hermitian Anderson models [7] and open quantum systems [8], to mention a few. While the implications of PT symmetry in the above mentioned fields are still debated, the theoretical demonstration and experimental observation of these concepts have been considered in the context of optics [9-16]. In linear domain, beam dynamics in such structures reveal that double refraction, power oscillations, and secondary emissions are possible [10,13]. By taking into account the effect of nonlinearity on beam dynamics, Z. H. Musslimani *et al* show that a novel class of one- and two-dimensional optical solitons can exist in optical PT lattices [11]. It's worth noting that PT optical potentials can be realized through a judicious inclusion of index guiding and gain/loss regions [9-16]. Given that the complex refractive-index distribution in guided wave geometry is $n(x) = n_R(x) + in_I(x)$, satisfying the PT condition $n_R(x) = n_R(-x)$ and $n_I(x) = -n_I(-x)$, one can deduce that $n(x)$ acts as an optical PT potential (where $x$ is the normalized transverse coordinate). The possibility of optical realization of relativistic non-Hermitian quantum mechanics, provided by the distributed-feedback optical structures with gain and/or loss regions has been proposed by S. Longhi [17].

One may note that all the PT potentials considered above are periodic. A natural question arises: how does light propagate if the PT potential has a local defect? Defect modes (DMs) in traditionally defective photonic lattice have been under consideration over the years [18-25]. One-dimensional linear and nonlinear DMs in 1D photonic lattice have been studied by F. Fedele *et al* [18] and J. Yang *et al* [19], respectively. While in the 2D geometry, linear and nonlinear localized DMs are systematically investigated by J. Wang *et al* [20] and W. Chen *et al* [21], respectively. The defect solitons in triangular and kagome optical latticess have also been reported [22-23]. Motivated by the above theoretical predictions, there are two typical experiments, one for 1D negative defect [24], and the other for 2D negative defect [25]. Quite recently, properties of linear and nonlinear defect modes in PT potentials have been investigated [26-27]. But the above two works are limited to defective PT potentials with real part of simple periodic lattices. Dual-frequency lattices constructed by spacing modulation have been demonstrated to support surface defect gap solitons without a threshold power [28]. In this contribution, we study the defect solitons supported by PT potentials whose real part is dual-frequency lattices with a single defect. The impact of defect on their stability properties is also explored.

## 2. Theoretical Modeling

To gain insight into the dynamics of soliton formation, we describe the propagation of light in a focusing nonlinear medium imprinted by PT potentials whose real part are dual-frequency lattices with a defect locating at the center, with the nonlinear Schrodinger equation for the dimensionless field amplitude $q$ [11]

$$iq_z + q_{xx} + |q|^2 q + V_0 [V(x) + iW(x)] q = 0. \tag{1}$$

Here the transverse $x$ and longitudinal $z$ coordinates are normalized by the input beam width $a$ and the diffraction length $L_{diff} = 2k_0 n_0 a^2$, respectively, where $k_0 = 2\pi/\lambda_0$ ($\lambda_0$ is the wavelength in vacuum) and $n_0$ is the constant background index. The depth of the PT potentials $V_0$ is scaled to $1/(2k_0^2 n_0 a^2)$. The function $V(x) + iW(x)$ represents the intensity profile of the PT potentials with one-site defect, the real part when $|x| \geq \pi/2$,

$$V(x) = \cos^2[\Omega_1(x+\pi/2)]\sin^2[\Omega_2(x+\pi/2)], \tag{2a}$$

and otherwise,

$$V(x) = (c+\varepsilon)\cos^2[\Omega_1(x+\pi/2)], \tag{2b}$$

while the imaginary part $W(x) = W_0 \sin(2x)$, where $\Omega_1$ and $\Omega_2$ describe the period and asymmetry of lattices (here and throughout the paper we let $\Omega_1 = 1$ and $\Omega_2 = 2$), $\varepsilon$ indicates the modulation depth of the peak intensity of the one-site defect, and $W_0$ characterizes the modulation depth of the peak intensity of the imaginary part of the PT potentials. Such refractive-index landscapes might be optically induced by an interference pattern in a photorefractive crystal using vectorial interactions [29]. We introduce the constant $c$ for the purpose of making the peak intensity of the defect without modulation ($\varepsilon = 0$) [Eq. (2a)] approximate to that of the real part of the PT potential [Eq. (2b)]. To investigate the impact of the lattice defect on the solitons, we set, unless stated otherwise, $c = 0.6$. A typical intensity profile of the PT potential is shown in Fig. 1(a).

We search for the stationary soliton solutions in the form of $q(x,z) = u(x)\exp(-ibz)$, where $u(x)$ is the complex function obeying the following nonlinear equation

$$u_{xx} + |u|^2 u + V_0[V(x) + iW(x)]u + bu = 0, \qquad (3)$$

and $b$ is the propagation constant. Eq. (1) conserves the total power (energy flow): $U = \int_{-\infty}^{+\infty} |u|^2 dx$.

Solving Eq. (2) by dint of the modified square-operator iteration method (MSOM) proposed by J. Yang [30], we can acquire the soliton profiles. Further, without loss of generality, we fix the depth of PT potentials by setting $V_0 = 3$, which is fixed unless another choice of $V_0$ is specified explicitly, and vary $b$ and $\varepsilon$.

In order to examine the linear stability, we then take perturbed solutions to Eq. (1) as $q(x,z) = \{u(x) + [v(x) - w(x)]\exp(\delta z) + [v(x) + w(x)]^* \exp(\delta^* z)\}\exp(-ibz)$, with perturbation eigenmodes $v$ and $w$ ($v, w \ll 1$), and instability growth rate $\delta$ (superscript "*" stands for the complex conjugation). Substitution of this expression in Eq. (1) and linearization around $u$ yields the eigenvalue problem

$$-i\begin{pmatrix} -iV_0W - i\,\mathrm{Im}(u^2) & L_0 + V_0V - \mathrm{Re}(u^2) \\ L_0 + V_0V + \mathrm{Re}(u^2) & -iV_0W + i\,\mathrm{Im}(u^2) \end{pmatrix}\begin{pmatrix} v \\ w \end{pmatrix} = \delta \begin{pmatrix} v \\ w \end{pmatrix}, \qquad (4a)$$

$$L_0 = \frac{d^2}{dx^2} + b + 2|u|^2, \qquad (4b)$$

which we solve, by adopting the original operator method (OOM) [31], to find perturbation profiles and associated growth rates $\delta$. The stability criterion of the system is that if $\mathrm{Re}(\delta) > 0$, the DSs are linearly unstable, and otherwise they are linearly stable [31].

## 3. Numerical simulations and discussion

Before investigating the existence and stability of the solitons, it is instructive to study the Floquet-Bloch spectrum of the nondefective PT potentials. The band-gap structure of the linear version of Eq. (1) with $V_0\left[\cos^2(x)\sin^2(2x)+iW_0\sin(2x)\right]$ is shown in Fig. 1(b), from which one may note that the boundaries and range of the semi-infinite (SI) gap and the first gap are $\mu<-0.783$ and $0.061<\mu<0.374$ for $W_0=0.05$, respectively. What's worth emphasizing is that there exists a critical value $W_0^{th}=0.1$, below which all the eigenvalues are real. Above the threshold, an abrupt phase transition occurs because of the spontaneous symmetry breaking and thus the partially complex band diagram forms. The above phenomena have analogy to the results proposed by K. G. Makris *et al* in periodic PT potentials [10-12].

Firstly we study the defect solitons whose real/imaginary parts are symmetric/antisymmetric functions of $x$, which were coined symmetric-antisymmetric defect solitons (SADSs). As three representative examples, $\varepsilon=-0.3$ for negative defects, $\varepsilon=0$ for no-modulation defect, and $\varepsilon=0.3$ for positive defects are under consideration. Fi2. (a)-(c) display the $U$ vs $b$ curves for $\varepsilon=-0.3$ (left column), $\varepsilon=0$ (middle column), and $\varepsilon=0.3$ (right column), respectively, from which we can find that for positive and no-modulation defects, the SADSs cease to exist at certain propagation constant in the SI gap (i.e., $b=-1.39$ for $\varepsilon=0.3$ and $b=-0.876$ for $\varepsilon=0$). For identical propagation constant, formation of SADSs demand more power for negative defects, but require less power for positive defects, compared to the no-modulation case. Similar to Ref. [19,27], this can be explained physically as follows: when the defect is lower than the surrounding lattice sites (negative defect), acting as a lower-index waveguide, light tend to escape from the defect site to the nearby sites, consequently more power are needed to trap the light. Conversely, positive defect can serve as a higher-index waveguide to guide light. Some characteristic SADS solutions are shown in Fig. 2 (d)-(i), from which we notice that the amplitude of real part of SADS increase as the $\varepsilon$ is lowered under the condition of same propagation constant [comparing Fig. 2 (d)-(f)]. While for a settled $\varepsilon$, following the drop in propagation constant, the profile of the real part of SADS becomes broader and the secondary maximum emerge, meanwhile the profile of the imaginary part of SADS becomes more twisted [comparing Fig. 2(d) and Fig. 2(g)].

A comprehensive stability analysis of SADSs showed that for positive and no-modulation defects, the SADSs are always stable in the SI gap and are nonexistent in the first gap, whereas for the negative defects, in the SI gap, the SADSs are stable in the most of their existence region except for those near the first band. Taking $\varepsilon = -0.3$ for example, as shown in Fig. 3(a) revealing the relationship between the real part of the growth rate $\mathrm{Re}(\delta)$ and propagation constant $b$, we found that the SADSs are stable for $b < -1.219$, which obey the VK criterion $dU/db < 0$ [32][Fig. 2(a)], but in the interval $[-1.219, 0-0.84]$ where $dU/db < 0$, the SADSs are unstable, violating the VK criterion. A point worth emphasizing is that for deep enough defects, in the first gap, the SADSs can be stable in the whole existence region. Choosing $\varepsilon = -0.55$ as an instance, we plot The power curve and a typical SADS solution of this case in Figure 3. It can be demonstrated that all the SADSs are stable in the whole existence region $-0.061 < b < 0.283$, which obey the VK criterion $dU/db < 0$ [Fig. 3(b)]. At the upper threshold of $b$, the SADSs cease to exist in the first gap.

To verify predictions of the above linear stability analysis, we solve the Eq. (1) with the input condition $q(x, z = 0) = u(x)[1 + \rho(x)]$ by employing the split-step Fourier method, where $\rho(x)$ is the random function with Gaussian distribution with its relative amplitude set at 10% level. Figure 4 presents some examples of stable and unstable propagation SADSs residing in both the SI gap and the first gap. In all the case studied (corresponding the examples marked by (a)-(d) in Figure 2 and Figure 3), the predictions of the linear stability analysis were confirmed. It was found that unstable SADSs were spitted in two filaments. In contrast, stable SADSs retain their structure indefinitely long, even in the presence of strong input noise.

Lastly, we investigate the defect solitons, the real/imaginary parts of which are antisymmetric/symmetric functions of $x$, which were named antisymmetric-symmetric defect solitons (ASDSs). To take $\varepsilon = 0.3$ for an example, we find that in the SI gap, the power of ASDSs firstly decreases with the growth of propagation constant and then exhibit an increasing behavior when approaching the edge of the first band, while the power of ASDSs in the first gap still decrease as a function of propagation constant[Fig. 5(a)]. Reprehensive examples of ASDS profiles are depicted in Fig. 5(c)-(e), from which one may note that in the SI gap, the real part of the ASDS is dipole function and the imaginary are even. The profile of the ASDS in the first gap exhibits a damped oscillation behavior and occupies many lattice sites. A huge of linear stability analysis and propagation simulation demonstrate that the ASDSs are unstable in the whole semi-infinite gap regardless of defects, but can be stable in the low power region near the second band for positive defects(i.e., $0.3 < b < 0.374$ for $\varepsilon = 0.3$) [fig. 5(b)]. The stable region of ASDS in first gap broadens as the modulation depth $\varepsilon$ becomes deep. The predictions of the above linear stability analysis are demonstrated by propagation simulations, as shown in Fig. 5(f)-(h) [corresponding to the fig. 5(c)-(e), respectively]. It was found that unstable ASDSs in the SI gap were spitted in two filaments. While unstable ASDSs in the first gap switch to neighbor sites after propagating a certain distance.

## 4. Conclusion

Summarily, we report on the existence and stability properties of DSs in parity-time potentials whose real parts are dual-frequency lattices with a defect locating at the center. The impact of defect on the stability regions of DSs was considered. For positive defects, SADSs are always stable in the SI gap and nonexistent in the first gap. While for negative defects, in semi-infinite gap, the SADSs are stable in most of their existence region apart from low power region, but in the first gap, all the SADSs are stable. The deeper the modulation depth becomes, the narrower the stable region of SADS in the SI gap is and the wider that in the first gap is. The ASDSs are unstable in the whole semi-infinite gap in spite of defects, but can be stable in the low power region for positive defects. The stable region of ASDS in first gap broadens as the modulation depth becomes deep.

**Acknowledgments**

This work was supported by the Scientific Research Foundation for the Returned Overseas Chinese Scholars, Specialized Research Fund for the Doctoral Program of Higher Education (200802871028), and the Summit of the "Six Great Talents" of Jiangsu Province (07-A-011)

**Figure captions**

Fig.1. (Color online) (a) The intensity profile without a modulation ($\varepsilon = 0$). Black solid: real part, dashed red: imaginary part. (b) Band gap spectrum of the PT potential, the real part of which is dual-frequency lattices ($V_0 = 3$, $W_0 = 0.05$).

Fig.2. (Color online) (a)-(c) Power diagrams of SADSs in the SI gap for $\varepsilon = -0.3$, $\varepsilon = 0$ and $\varepsilon = 0.3$, respectively (solid lines: stable; dashed lines: unstable). (d)-(f) Profiles of SADSs with $b = -6$ in the SI gap [marked by circles in (a)-(c)]. (g) Profile of SADS with $b = -0.95$ for $\varepsilon = -0.3$. (h) Profile of SADS with $b = -1$ for $\varepsilon = 0$. (i) Profile of SADS with $b = -1.45$ for $\varepsilon = 0.3$ [marked by triangles in (a)-(c)]. Shaded part in (a)-(c): Bloch band.

Fig.3. (Color online) (a) Dependence of $\text{Re}(\delta)$ on $b$ of SADSs in the SI gap for $\varepsilon = -0.3$. (b) Power diagram of SADS in the first gap for $\varepsilon = -0.55$. (c) Profile of SADS with $b = 0.15$ for $\varepsilon = -0.55$ [marked by circle in (b)]. Shaded part in (a)-(b): Bloch bands.

Fig.4. (Color online) Stable and unstable propagation of SADSs in both gaps. $b = -0.95$ for $\varepsilon = -0.3$ in (a), $b = -1$ for $\varepsilon = 0$ in (b), $b = -6$ for $\varepsilon = 0.3$ in (c), and $b = 0.15$ for $\varepsilon = -0.55$ in (d), corresponding a-d marked in Fig. 2 and Fig. 3, respectively.

Fig.5. (Color online) (a) Power diagram of ASDSs in both gaps for $\varepsilon = 0.3$. (b) Instability growth rate of ASDSs. (c)-(e) Profiles of three ASDSs at $b = -1.2, 0.1$, and $0.32$ [marked by circle, square, and triangle], respectively. (f)-(h) Stable and unstable propagation of ASDSs in both gaps, corresponding to (c)-(e), respectively. Shaded part in (a): Bloch bands.

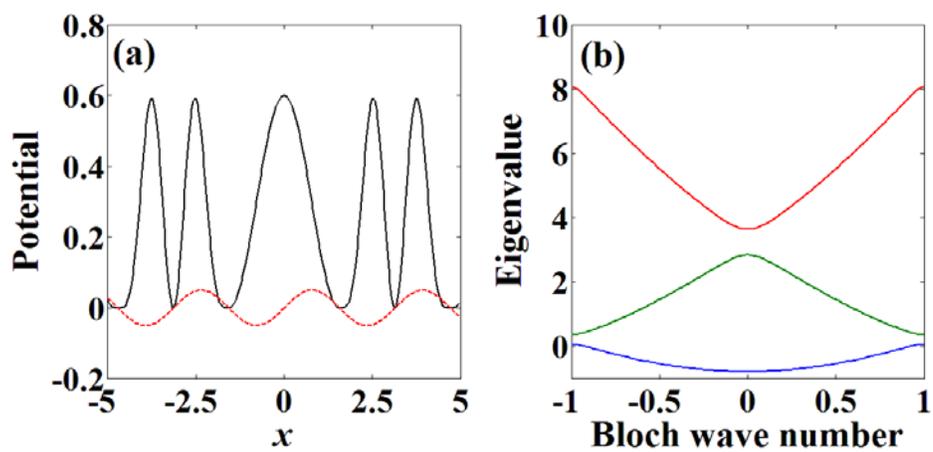

**Fig. 1 (by Meng *et al.*)**

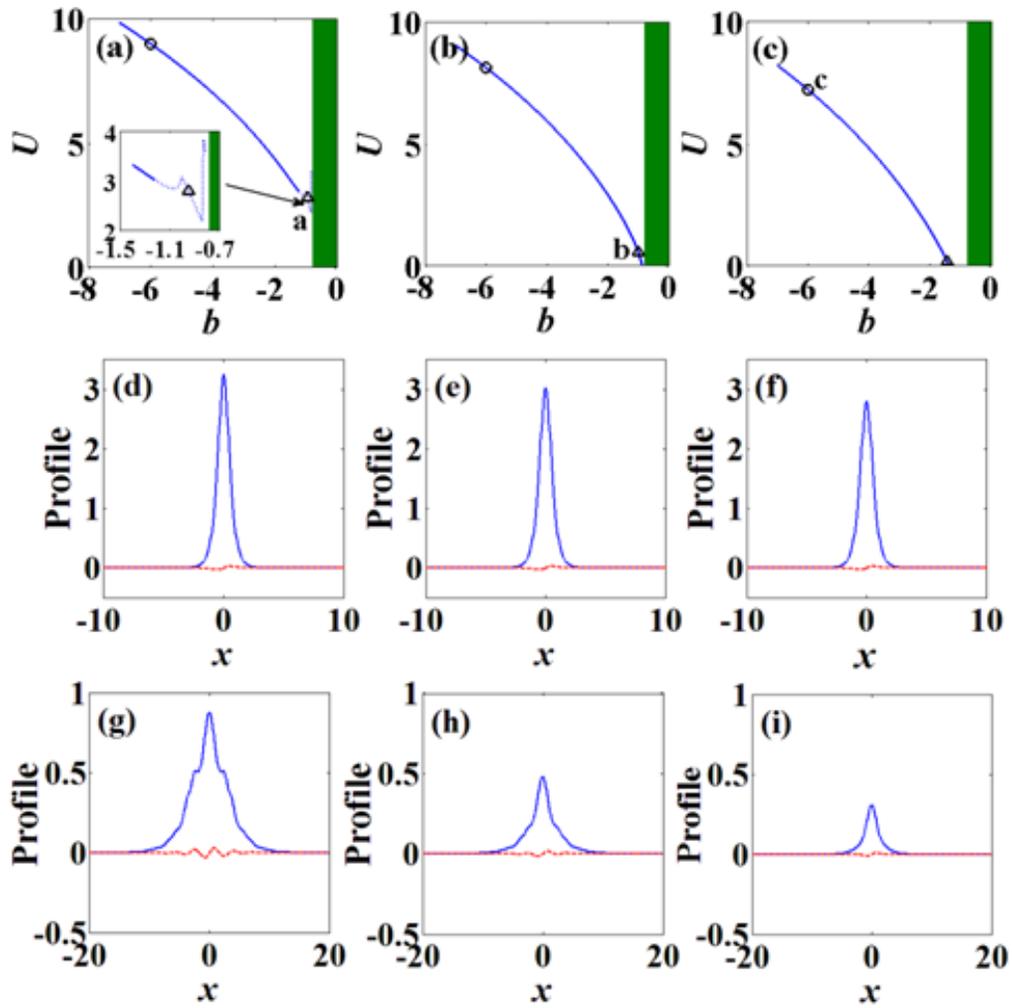

**Fig. 2 (by Meng *et al.*)**

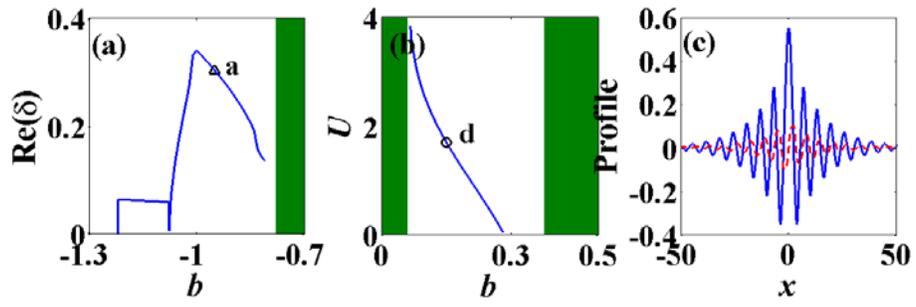

**Fig. 3 (by Meng *et al.*)**

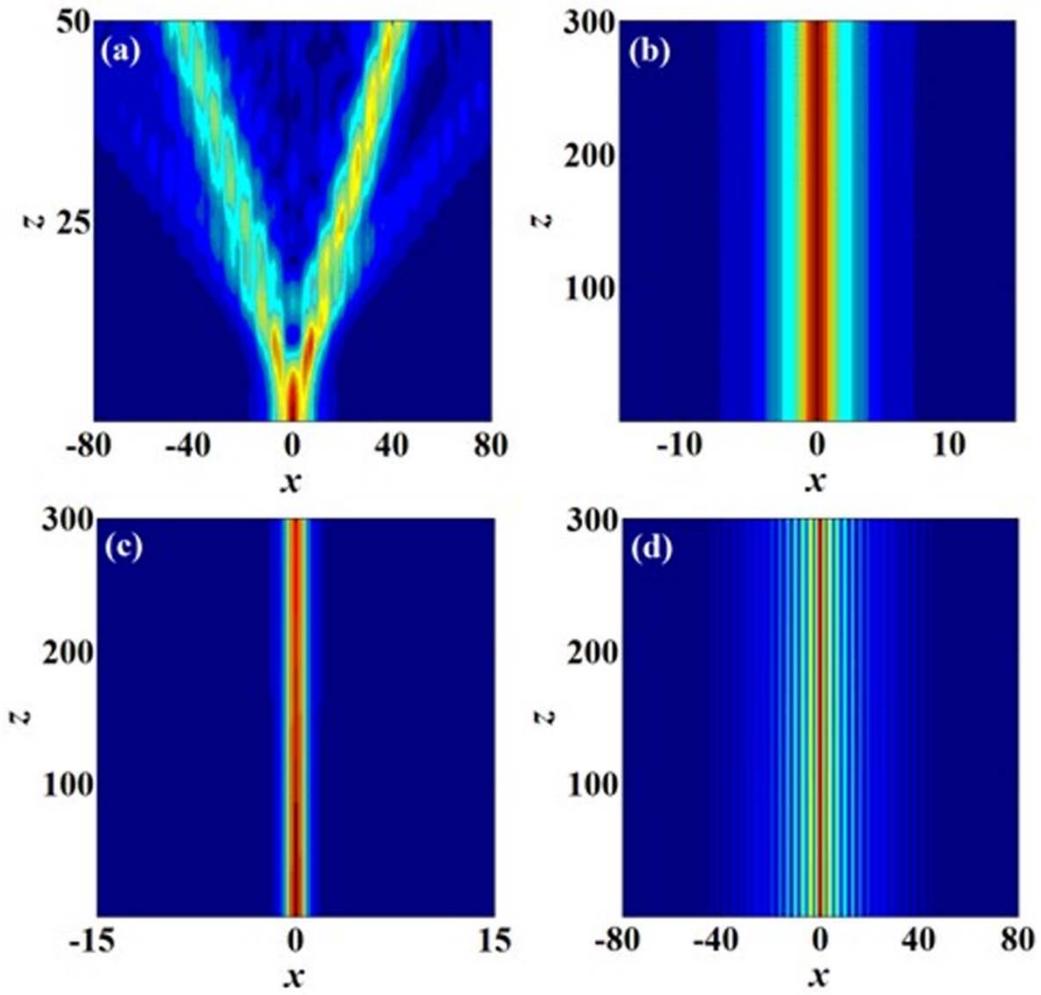

**Fig. 4 (by Meng *et al.*)**

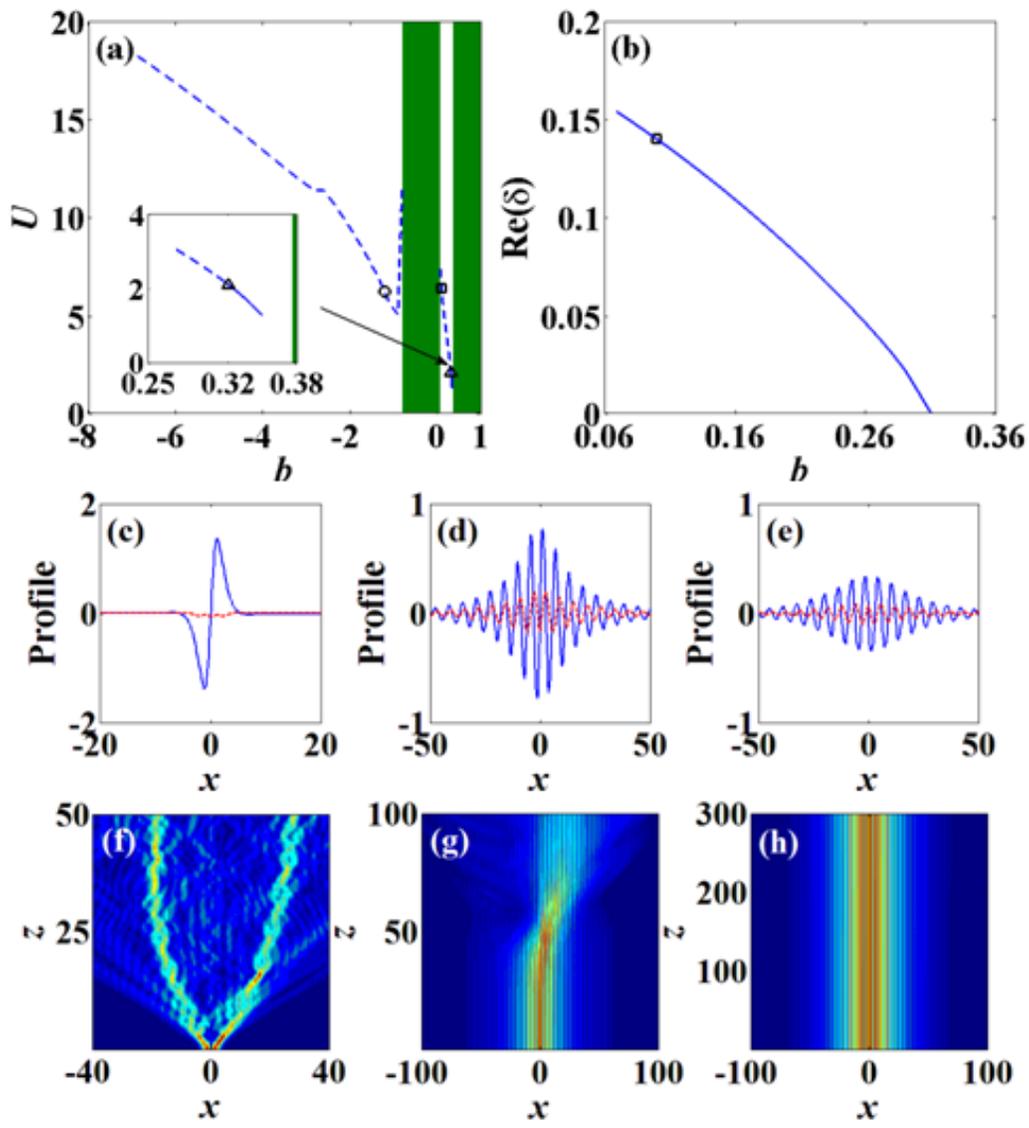

**Fig. 5 (by Meng *et al.*)**